\documentclass[submission, Proceedings,a4paper]{SciPost}

\usepackage{booktabs}
\usepackage{colortbl}
\definecolor{myblue}{rgb}{0.8,0.85,1} 
\definecolor{light-gray}{gray}{0.95}

\def\beq {\begin{equation}}
\def\eeq {\end{equation}}
\def\bea {\begin{eqnarray}}
\def\eea {\end{eqnarray}}

\def\Dlog{{\rm Dlog}}

\newcommand{\MSb}{{\overline{\rm MS}}}
\newcommand{\wh}{\widehat}

\newcommand{\ket}[1]{\left|#1\right\rangle}

\begin{document}

\begin{center}{\Large \textbf{\boldmath
 Renormalization-group improvement  in
 hadronic $\tau$ decays in  2018 
      }}\end{center}

\begin{center}
D. Boito\textsuperscript{1*},
P. Masjuan\textsuperscript{2},
F. Oliani\textsuperscript{1}
\end{center}

\begin{center}
{\bf 1} Instituto de F\'isica de S\~ao Carlos, Universidade de S\~ao Paulo, CP 369, 13560-970, S\~ao
Carlos, SP, Brazil
\\
{\bf 2} Grup de F\'{\i}sica Te\`orica, Departament de F\'{\i}sica, 
  Universitat Aut\`onoma de Barcelona,  and Institut de F\'{\i}sica d'Altes Energies (IFAE), 
  The Barcelona Institute of Science and Technology (BIST), 
  Campus UAB, E-08193 Bellaterra (Barcelona), Spain
\\
* boito@ifsc.usp.br
\end{center}

\begin{center}
\today
\end{center}

\definecolor{palegray}{gray}{0.95}
\begin{center}
\colorbox{palegray}{
  \begin{tabular}{rr}
  \begin{minipage}{0.05\textwidth}
    \includegraphics[width=8mm]{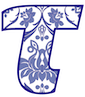}
  \end{minipage}
  &
  \begin{minipage}{0.82\textwidth}
    \begin{center}
    {\it Proceedings for the 15th International Workshop on Tau Lepton Physics,}\\
    {\it Amsterdam, The Netherlands, 24-28 September 2018} \\
    \href{https://scipost.org/SciPostPhysProc.1}{\small \sf scipost.org/SciPostPhysProc.Tau2018}\\
    \end{center}
  \end{minipage}
\end{tabular}
}
\end{center}


\section*{Abstract}
{\bf
One of the main sources of theoretical uncertainty in the extraction of the strong coupling 
from hadronic tau decays stems from the renormalization group improvement of the series. Perturbative 
series in QCD are divergent but are (most likely) asymptotic expansions.  One needs knowledge 
about higher orders to be able to  choose the optimal renormalization-scale setting procedure. 
Here, we discuss the use of Pad\'e approximants
as a model-independent and robust method to extract information about the higher-order terms. We show that 
in hadronic \boldmath $\tau$ decays
the fixed-order expansion, known as fixed-order perturbation theory (FOPT), is the most reliable mainstream 
method to set the scale. 
This fully corroborates previous conclusions based on the available knowledge about the leading renormalon 
singularities of the perturbative series. 
 }

\vspace{10pt}
\noindent\rule{\textwidth}{1pt}
\tableofcontents\thispagestyle{fancy}
\noindent\rule{\textwidth}{1pt}
\vspace{10pt}

\section{Introduction}
\label{sec:intro}

Since the 1990s, inclusive hadronic decays of the $\tau$ lepton have
been acknowledged as a reliable source of information about QCD. In
particular, the strong coupling, $\alpha_s$, can be extracted with
competitive precision from these decays. Since the works by Braaten,
Narison and Pich~\cite{BNP92} and later by Le Diberder and
Pich~\cite{LDP92}, which finally shaped the standard strategy to
extract $\alpha_s$ from this process, several important developments
have occured. On the experimental side, the precision has improved a
lot thanks to the LEP experiments; the latest (re)analysis of ALEPH
data was published in 2014~\cite{ALEPH}. On the theory side, our
understanding of the theoretical input from QCD necessary to achieve
an accurate $\alpha_s$ determination has improved as well.  In
parallel, there was similar progress in the global knowledge about
$\alpha_s$ from other processes in the past 25 years. The uncertainty
in the PDG recommendation for $\alpha_s(m_Z)$ went down from about
$5\%$ in 1994~\cite{PDG94} to a mere $0.9\%$ in the latest
edition~\cite{PDG18,GS17}, while individual extractions from the
lattice are achieving uncertainties below $1\%$ (see, for example,
\cite{ALPHACollab}).  Although the extraction of $\alpha_s$ from
$\tau$ decays remains appealing --- it is performed at rather
low-energies and provides, therefore, a non-trivial test of asymptotic
freedom --- it must be carefully scrutinized given the state of
affairs.

In the last few years, a reassessment of the $\alpha_s$ extraction from
$\tau$ decays was motivated by the publication of the result for the
$\alpha_s^4$ correction in the relevant perturbative QCD series, which
is the next-to-next-to-next-to-leading order (N3LO) correction~\cite{BCK08,NIKHEF}. This
tour de force calculation, a five-loop QCD result involving about
20,000 Feynman diagrams, was completed in 2008 --- more
than 15 years after the publication of the $\alpha_s^3$
result~\cite{alphas3ref1,alphas3ref2}.  Since then, many aspects of
the extraction of $\alpha_s$ from $\tau$ decays have been
reexamined. In this note, we will focus on the perturbative series, in
particular on its  renormalization group improvement.  

The QCD description of
hadronic tau decays must rely on finite-energy sum rules, which
exploit analyticity in order to circumvent the breakdown of
perturbative QCD at low energies. The theoretical predictions are then
obtained from a contour integral in the complex plane of the variable
$s$ --- the invariant mass of the final-state hadrons. When performing
this integration, one must set the renormalization scale. The two most
common procedures are known as fixed-order perturbation theory (FOPT),
in which the scale is kept fixed, and contour-improved perturbation
theory (CIPT)~\cite{LDP92,Pivovarov}, in which the scale runs along the contour of
integration. The two lead to different series and this difference, which is larger than the error
ascribed to each series  individually, is
one of the main sources of theoretical uncertainty in the extraction
of $\alpha_s$ from hadronic $\tau$ decay data.

Before entering the specifics of $\tau$ decays let us remind some
basic facts about perturbative expansions in QCD. As discovered by
Dyson in 1952, the perturbative series in powers of the coupling in
realistic quantum field theories are divergent expansions~\cite{Dyson},
no matter how small the coupling is. The fact that the first few terms
of these series do provide meaningful results, i.e. they seem to agree
reasonably well with experiment, led Dyson to conjecture that these
series must be asymptotic expansions: a special type of divergent
series that are useful in practice. Asymptotic expansions approach the
true value of the function being expanded up to a finite order, after which
the series starts to diverge. 
Their usefulness 
is illustrated by the famous Carrier's rule which states that
\begin{quotation}
``Divergent
series converge faster than convergent series because they don't have
to converge''~\cite{Boyd}.
\end{quotation}
The idea is that the series may approach
the true value much faster than a convergent expansion, which is
actually a fortunate feature in QCD, since the computation of
higher-order corrections becomes quickly impractical. It also implies
that a good asymptotic expansion, in  comparison with a convergent one,
can have consecutive terms that decrease less in magnitude when compared to their predecessor, precisely
because ``it does not have to converge''.

The divergence of the series is due to the factorial growth of its
coefficients --- this behaviour, in turn, can be mapped to regions
of specific loop diagrams. The ``optimal truncation'' of an asymptotic
series of this type is often achieved by truncating it at its smallest
term. This goes under the name superasymptotic approximation~\cite{Boyd}.
The error that is made in such an approximation is typically of the
order of $e^{-p/\alpha}$, where $p>0$ is a constant and $\alpha$ the
expansion parameter.  The quantity $e^{-p/\alpha}$ is non-perturbative
and does not admit a power series in $\alpha$, but vanishes when the
expansion parameter goes to zero, as one would intuitively
expect.\footnote{Given the logarithmic running of $\alpha_s$ in QCD the 
error of the truncated perturbative QCD expansion becomes 
$e^{-p/\alpha_s(Q)} \sim \left(\frac{\Lambda^2}{Q^2}  \right)^p$.
These non-perturbative power corrections in $1/Q^2$ are, of course,
related to the higher-order terms in the Wilson's OPE. In the case of
$\tau$ decays, the leading one is given by the gluon condensate and
scales as $1/Q^4$. There is an infinite series of such terms, one for
each gauge-invariant operator that contributes to the OPE and,
therefore, the QCD expansion becomes a double expansion in $\alpha_s$
and in $1/Q^2$~\cite{Renormalons}. Effects related to asymptotic nature 
of the latter are related to the so-called duality 
violations~\cite{DVs2018,PerisProceedings}.} This suggests that the issues related to the fact that
the series is asymptotic become more prominent when the coupling is
larger. In QCD this means lower energies, such as in $\tau$ decays
where the relevant scale, of the order of the $\tau$ mass is $\sim
2$~GeV.  One should say that
these rules do not have the status of theorems, mathematical proofs
are rare  here. As a matter of fact,
there is no proof that the series in QCD is asymptotic to start with,
but everything indicates that this is indeed the case.

The discussion about FOPT and CIPT in $\tau$ decays and a decision
about which one is the most reliable procedure cannot be taken out of
this context.  They are both asymptotic series (at best), therefore
divergent.  In particular, some of the arguments put forward in the
literature in favour of CIPT~\cite{PichTau}\footnote{This argument~\cite{PichTau} 
is essentially unaltered since the publication of Ref.~\cite{LDP92} in 1992.} based on the relative size
of the first few terms of the series are insufficient, since they tend
to downplay, or simply ignore, these basic features of perturbative
expansions in QCD.  Also, in analysing the series  it does not make sense to talk about a
radius of convergence, because we are dealing with divergent expansions.
Inevitably, a final conclusion about the reliability of the two
different procedures requires knowledge about higher-order
coefficients of the series. In the absence of higher-order loop
computations, one must resort to other methods to estimate those terms.

A possibility is to use our partial knowledge about the renormalons of
the series. It is well known that the behaviour of a series of this
type at intermediate and high orders is dominated by the renormalons
close to the origin. Under reasonable assumptions, one can then
construct an approximation to the Borel transform of the series\footnote{Essentially its inverse Laplace transform.} using
the leading renormalons and match this description to the exactly know
coefficients. This allows for an extrapolation to higher orders and
one is able to obtain an estimate for the higher-order coefficients.
These type of construction has been studied in detail in
Refs.~\cite{BJ08,BBJ13}. The main conclusions that can be drawn from
this strategy are twofold.
\begin{enumerate}
  \item Under reasonable assumptions, i.e.,
without any artificial suppression of leading renormalon singularities,
FOPT is the most reliable method to set the renormalization scale in
hadronic $\tau$ decays. Because in CIPT a subset of terms, associated
with the running of the coupling, are resummed to all orders important
cancellations are missed and the series does not provide a good
approximation to the ``true'' value --- understood as the value
obtained from the Borel sum of the reconstructed series.
\item The
fact that FOPT is to be preferred is linked to the renormalon
singularity associated with the gluon condensate. Should this
singularity be, for some unknown reason, much suppressed then CIPT
would be best.
\end{enumerate}
These conclusions, albeit providing strong support to FOPT, are
somewhat model dependent since they do rely on the partial knowledge
about the renormalons and could be affected by the inclusion or the
removal of a specific singularity from the model. It is, therefore,
desirable to study this issue from a model-independent point of view
in order to corroborate, or to discredit, the results obtained from
the renormalon models.

Here we will discuss recent results presented in Ref.~\cite{BMO18}
where we used the mathematical method of Pad\'e
approximants~\cite{Baker} to extract information about the
higher-order coefficients of the series. Pad\'e, or rational,
approximants are a reliable model-independent tool that has regained
importance in recent years and has found applications in many aspects
of particle physics~\cite{MP08,MP07,WSDB18,CMRS16,gm2}.  In Ref.~\cite{BMO18} we applied
the method systematically to the problem of estimating higher orders
in the perturbative QCD description of hadronic $\tau$ decays. We
first used the large-$\beta_0$ limit of QCD where the series is
exactly known to all orders in $\alpha_s$ to test the method.  This
was done having in mind the concrete situation of QCD, i.e.,
reconstructing the series solely from its first four coefficients.
The method has proven to be robust and sufficiently precise to allow
for a conclusion about the reliability of FOPT and CIPT, correctly
reproducing the fact that FOPT is to be preferred in the
large-$\beta_0$ limit. We then turned to QCD and applying the same
methods reconstructed the higher orders of the series. Our main conclusions were \begin{itemize}
  
\item The results from  Pad\'e approximants  and its variants are robust.
  This conclusion is supported both by the tests in
  large-$\beta_0$ and by the fact that we are able to obtain the N3LO
  coefficient in QCD from the lower order ones with good precision.

\item The reconstruction based on the model-independent Pad\'e
  approximants favours FOPT and lends support to the renormalon models
  of Refs.~\cite{BJ08,BBJ13}.

\item The six-loop coefficient of the Adler function is found to be $c_{5,1}=277\pm 51$. This result is in line with some other estimates~\cite{BJ08,BCK02}, but has a smaller uncertainty.
  \end{itemize}

In the remainder of this note, we will review the main results of Ref.~\cite{BMO18} to which we refer for further details.

\section{Overview of the theory}

\subsection{QCD in hadronic $\tau$ decays}
Here, we briefly recall the main theoretical ingredients needed for
the QCD analysis of hadronic $\tau$ decays. We refer to
Refs.~\cite{BMO18,BJ08,BBJ13} for further details.

The main observable in hadronic $\tau$ decays is the ratio $R_\tau$
which represents the total decay width normalized to the width of
$\tau \to e \bar \nu_e \nu_\tau$. Here, we restrict the analysis to non-strange
channels which allows us to safely neglect effects due to quark masses.
There are then two observables $R_{\tau,V}$ and $R_{\tau,A}$ where the decay
is mediated by vector and axial-vector $\bar u d$ currents, respectively. They
can be parametrized as
\beq
R_{\tau,V/A} = \frac{N_c}{2}S_{\rm EW} |V_{ud}|^2 \left[1 + \delta^{(0)} + \delta_{\rm NP}  +\delta_{\rm EW}\right],
\eeq
where $S_{\rm EW}$ and $\delta_{\rm EW}$ are small electroweak
corrections and $V_{ud}$ the CKM matrix element, $\delta_{\rm NP}$
encloses all non-perturbative corrections both from OPE condensates
and from duality-violations. The unity in between square brackets is
the partonic result while $\delta^{(0)}$, which is the main object of
this work, represents the perturbative QCD corrections.

The relevant quark-current correlators are
\beq
\Pi_{V/A}^{\mu \nu}(p) \equiv i \int dx \,e^{i p x}\, \langle \Omega | T \{ J_{V/A}^{\mu}(x) J_{V/A}^{\nu}(0)^{\dagger} \} | \Omega \rangle,\label{eq:Corr}
\eeq
where $\ket{\Omega}$ represents the physical vacuum and the currents
are $J^\mu_{V/(A)}(x)=(\bar u \gamma^\mu (\gamma_5)d)(x)$. They admit the usual
decomposition into transverse $\Pi^{(1)}_{V/A}(s)$,
and longitudinal, $\Pi^{(0)}_{V/A}(s)$, parts. Because the correlators depend
on conventions related to the renormalization procedure, it is advantageous to work with the Adler function, defined as the logrithmic derivative of  $\Pi^{(1+0)}(s)$ as 
$D^{(1+0)}(s)= -s \frac{d}{ds}\left[ \Pi^{(1+0)}(s)\right].$
Exploiting the analyticity of the correlators involved, the
perturbative corrections are written as integral in the complex plane
with fixed $|s|=m_\tau^2$ as~\cite{BJ08}
\beq
 \delta^{(0)} = \frac{1}{2\pi i}  \oint\displaylimits_{|x|=1}\frac{dx}{x} W(x) \widehat D^{(1+0)}_{\rm pert}(m_\tau^2x),\label{eq:delta0}
 \eeq
 where $x=s/m_\tau^2$ and $W(x)$ is the weight function determined by kinematics.
The perturbative expansion of $\widehat D$  starts at $\mathcal{O}(\alpha_s)$ and can be cast as 
\beq
\widehat D_{\rm pert}(s) =  \sum\limits_{n = 1}^{\infty}{a^{n}_{\mu}} \sum\limits_{k = 1}^{n+1 }{k c_{n,k}L^{k-1}},\label{eq:AdlerExp}
\eeq
where $L=\log(-s/\mu^2)$ and $a_\mu =\alpha_s(\mu)/\pi$.
In this
expansion, the only independent coefficients are the $c_{n,1}$; the
others can be obtained imposing renormalization group (RG) invariance,
and are expressed in terms of the $c_{n,1}$ and $\beta$-function
coefficients~\cite{BJ08,MJ05}. The logarithms  can be summed with the scale choice $\mu^2 = -s \equiv Q^2$ giving
\beq
\widehat D(Q^{2}) =   \sum_{n=0}^\infty r_n \alpha_s^{n+1}(Q)\equiv  \sum_{n=0}^\infty c_n a_s^{n+1}(Q) = a_Q + 1.640 \, a_Q^2 +6.371\, a_Q^3+ 49.08\, a_Q^4+\cdots,\label{eq:DCIPT}
\eeq
where $r_n=c_{n+1,1}/\pi^{n+1}$ and the numerical coefficients correspond to the choice $\mu^2=Q^2$,  $N_f=3$, in the $\MSb$ scheme.

To obtain the perturbative corrections to $R_{\tau,{V/A}}$ one needs to perform the integral in Eq.~(\ref{eq:delta0}). In the process,
one  needs to adopt a procedure in order to set the scale $\mu$, which
enters, implicitly, through Eq.~(\ref{eq:AdlerExp}). A
running scale, $\mu^2=Q^2$, as done in Eq.~(\ref{eq:DCIPT}), gives rise to
the aforementioned Contour-Improved Perturbation Theory (CIPT), where the running of
$\alpha_s$ along the contour is resummed to all orders. In this case, $\delta^{(0)}$ can
be written as
\beq
\delta^{(0)}_{\rm CI} = \sum_{n=1}^{\infty} c_{n,1}J_n^{\rm CI}(m_\tau^2), \qquad \mbox{with}  \qquad J_n^{\rm CI}(m_\tau^2)=  \frac{1}{2\pi i}\oint\displaylimits_{|x|=1} \frac{dx}{x}(1-x)^{3}(1+x)a^{n}(-m_\tau^2x).
\eeq
Another option is to employ a fixed scale
$\mu^2=m_\tau^2$, which gives rise to  Fixed Order Perturbation
Theory\footnote{Here we will consider only CIPT and FOPT, but alternative schemes for setting the scale $\mu$ have been advocated in the literature~\cite{Caprinietal1,Caprinietal3,Caprinietal4,CLMV10}.}. Then, because $\alpha_s$ is evaluated at a fixed
scale, it can be taken outside the contour integrals, which are performed over
 the logarithms that appear in Eq.~(\ref{eq:AdlerExp}) as
\beq
\delta^{(0)}_{\rm FO} = 	\sum\limits_{n=1}^{\infty}{a_{\tau}^{n}} \sum\limits_{k=1}^{n}{k c_{n,k}} J^{\rm FO}_{k-1},\qquad \mbox{with} \qquad J_{n}^{\rm FO} \equiv \frac{1}{2 \pi i}\oint\limits_{|x|=1}{\frac{dx}{x}(1-x)^{3}(1+x)\ln^{n}(-x)}.\label{FOPTdef}
\eeq
Therefore, $\delta^{(0)}_{\rm FO}$ can also be written as an expansion in the coupling
where the coefficients depend then on $c_{n,1}$, on the $\beta$-function coefficients, and on the integrals $J^{\rm FO}_n$. In QCD, this expansion reads, for $N_f=3$ and in the $\MSb$ scheme,
\beq
\delta^{(0)}_{\rm FO} =  \sum_{n=1}^\infty d_n a^n_Q=a_Q + 5.202\,  a_Q^2 + 26.37\, a_Q^3 + 127.1 \,a_Q^4 +( 307.8+c_{5,1})\,a_Q^5 + \cdots\label{FOPTexp}
\eeq
where we give the numerical result of the known contributions to the first unknown coefficient.

Because the perturbative series is divergent, it is convenient to work with the Borel transformed series, which can have  a finite radius of convergence, defined, in terms of the expansion in $\alpha_s$,  as
\beq
B[\widehat R](t) \equiv \sum\limits_{n=0}^{\infty}{r_{n} \frac{t^{n}}{n !}}.
\label{BorelDef}
\eeq
The original expansion  can be understood as an asymptotic series to the inverse Borel transform
\beq
\widehat R (\alpha) \equiv \int\limits_{0}^{\infty}{dt \text{e}^{-t/\alpha}B[\widehat R](t)},\label{BorelInt}
\eeq
provided  the integral exists. In our context,  the series $\widehat R$ can represent either the reduced Adler function, $\widehat D$ of Eq.~(\ref{eq:DCIPT}), or $\delta_{\rm FO}^{(0)}$ given in Eq.(\ref{FOPTexp}). The last equation defines the Borel sum of the asymptotic series. The divergence of the original series, $\widehat D$, is translated into singularities in the $t$ variable. Two types can be
distinguished: ultraviolet (UV) and infrared (IR) renormalons.
 The UV renormalons lie on the negative real axis and contribute
with sign alternating coefficients. IR renormalons are singularities on the positive real axis which contribute with
fixed sign coefficients. The latter  obstruct the integration in Eq.~(\ref{BorelInt}) and generate an
ambiguity in the inverse Borel transform which is expected to cancel against power corrections of the OPE. The position of the singularities in the $t$ plane can be determined with general renormalization group (RG) arguments. For the Adler function,  they appear at
positive and negative integer values of the variable $u\equiv \frac{\beta_1 t}{2\pi}$ (except for $u=1$), where $\beta_1$ is the leading coefficient of the QCD $\beta$-function.\footnote{We define the QCD $\beta$-function as 
\[
\beta(a_\mu) \equiv -\mu \frac{d a_\mu}{d\mu} = \beta_1a_\mu^2 +\beta_2a_\mu^3+\beta_3a_\mu^4 + \beta_4a_\mu^5+\beta_5a_\mu^6+\cdots
\]}  The UV renormalon at
$u=-1$, being the closest to the origin, dominates the large order behaviour of the series, which must, therefore, be sign alternating at  higher orders. As seen in Eq.~(\ref{eq:DCIPT}), this sign alternation is still not apparent   in the first four coefficients of the QCD expansion in the $\MSb$ scheme, which are known exactly.

\subsection{Pad\'e approximants}

A Pad\'e approximant (PA) to a function $f(z)$~\cite{Baker}, denoted $P^M_N(z)$, is defined as the ratio of two polynomials in the variable $z$ of order $M$ and $N$, $Q_M(z)$ and $R_N(z)$, respectively, with the definition $R_N(0)=1$. Let us consider a function $f(z)$  which assumes a series expansion around $z=0$ as 
\beq
f(z) = \sum_{n=0}^{\infty} f_n z^n.\label{fz}
\eeq
The Pad\'e $P_N^M(z)$ is said to have a ``contact" of order $M+N$ with the expansion of the function $f(z)$ around the origin of the complex plane: the expansion of $P^M_N(z)$ around the origin  is the same as that of $f(z)$ for the first $M+N+1$ coefficients
\beq
P_N^M(z) = \frac{Q_M(z)}{R_N(z)} \approx f_0 +f_1\,z+f_2\,z^2+\cdots+ f_{M+N}z^{N+M} + \mathcal{O}\left( z^{M+N+1}  \right).
\eeq
From the reexpansion
of the approximant $P_N^M(z)$ one can read off an estimate for the
coefficient $f_{M+N+1}$, the first that is not used as input~\cite{MP08}. Estimates of this type will be of special
interest in this work.

The successful use of Pad\'e approximants to obtain quantitative
results about the function $f(z)$ requires only a qualitative
knowledge about the  analytic properties of the function. The PAs
can also be used to perform a reconstruction of the singularity
structure of $f(z)$ from its Taylor expansion. Convergence theorems
exist for the cases of analytic and single-valued functions with
multipoles or essential singularities~\cite{Baker}. Even for functions that
have branch points the PAs can be used, in many cases,
successfully. In these cases, for increasing order of approximation,
the poles of the PAs tend to accumulate along the branch cut,
effectively mimicking the analytic structure of the function~\cite{Baker}.

In this work, most of the times, the role of the function $f(z)$ is played  by the
Borel transform of the Adler function, defined in
Eq.~(\ref{BorelDef}). A key feature of the Borel transform, as already
discussed, is its singularities along the real axis, the
renormalons. It will be of interest to us to study how this
singularity structure is mimicked by the PAs. It is important to note that
when $f(z)$ is a general
meromorphic function some of the poles (and residues) of the
approximant $P_N^M(z)$ may become complex, even though the original
function has no complex poles.\footnote{When the meromorphic function is of the Stieltjes type the poles will always be  along the real axis. The functions we approximate in this work are not of this type. } Such poles cannot be identified with
any of the renormalon singularities, but they do not prevent the use
of $P_N^M(z)$ to study the function away from these poles. In fact, in the
process of approximating a function with an infinite number of poles
by an approximant that contains only a handful of them, the appearance of
these extraneous poles is expected to
happen~\cite{MP07}.

The approximation of functions with branch points and cuts --- as is
the case for the Borel transform of the Adler function in QCD --- is
more subtle. In this case, a possible strategy is  the
manipulation of the series to a form which is more amenable to the  approximation by  Pad\'es. Let us consider the particular case of a function $f(z) =
\frac{A(z)}{(\mu-z)^{\gamma}} + B(z)$ with a cut from $\mu$ to
$\infty$ with exponent $\gamma$ and a reminder $B(z)$ with little
structure (both $A(z)$ and $B(z)$ are to be analytic at $z=\mu$). Following the method of Baker called D-log Pad\'e
approximant~\cite{Baker}, we can form PAs not to
$f(z)$ but to
\begin{equation}\label{Dlog}
F(z) = \frac{\rm d}{ {\rm d}z} \log[f(z)] \sim \frac{\gamma}{\mu-z} \quad \quad (\textrm{near } z=\mu)\, ,
\end{equation}
which turns out to be a meromorphic function to which the convergence theorem applies.
The use of appropriate Pad\'e approximants to $F(z)$ determines
in an unbiased way
both  the pole position, $z=\mu$, and the residue, $-\gamma$, which
  corresponds to the exponent of the cut of $f(z)$.
No assumption about neither $\mu$ nor $\gamma$ is made; they are determined
directly from the series coefficients. The approximation of $F(z)$ by a PA
yields an approximant for $f(z)$ that is not necessarily a rational function. To be more specific, the Dlog-PA approximant  to $f(z)$ obtained from using $P_N^M$ to approximate $F(z)$, that we denote $ {\rm Dlog}_N^M(z)$,  is 
\begin{equation}\label{DlogNM}
{\rm Dlog}_{ N}^M(z) = f(0) e^{\int d z \frac{Q_M(z)}{R_N(z)}}\, ,
\end{equation}
\noindent
where $P_N^M(z)= \frac{Q_M(z)}{R_N(z)} $ is the aforementioned PA to
$F(z)$. Due to the derivative in Eq.~(\ref{Dlog}), the constant
$f(0)$ is lost and must be reintroduced in order to properly normalize the
${\rm Dlog}^M_N(z)$. In practice, the non-rational approximant ${\rm Dlog}_{ N}^M(z)$ can  yield a rich analytical structure, in particular the presence of
branch cuts --- not necessarily present in the function $f(z)$ --- is to be expected.

\section{\boldmath Results in large-$\beta_0$}

Before discussing our results in QCD, we will present  results in the so-called large-$\beta_0$ limit, which is a
 good laboratory for the strategy we present here. Results in this limit are obtained
by first considering a large number of fermion flavours, $N_f$,
keeping $\alpha_s N_f\sim1$. In this framework, the $q\bar q$ bubble
corrections to the gluon propagator must be resummed to all orders.
Using  this dressed gluon propagator
one can then compute all the corrections with highest power of  $N_f$  at every $\alpha_s$ order to  a given QCD
observable~\cite{Renormalons}.
The results in large-$\beta_0$ are obtained by replacing the $N_f$
dependence by the leading QCD $\beta$-function coefficient ($\beta_1$ in our notation) which incorporates a set of non-abelian gluon-loop diagrams.
 Accordingly,
 the QCD $\beta$-function is truncated at its first term.\footnote{Strictly speaking,  the large-$\beta_0$ limit would  be the ``large-$\beta_1$'' limit, in our notation. }

 In this limit, the Borel transform of the reduced Adler function, defined in Eq.~(\ref{BorelDef}) can be written in a closed form  as~\cite{MB93,DB93,Renormalons}
\beq
B [\widehat D](u) = \frac{32}{3 \pi} \frac{e^{(C+5/3) u}}{(2-u)} \sum\limits_{k=2}^\infty{\frac{(-1)^k k}{[k^2-(1-u)^2]^2}},\label{BDLb}
\eeq
where the scheme parameter $C$ measures the departure from the $\MSb$, which
corresponds to the choice $C=0$. The result clearly exhibits the renormalon poles, both the
IR, that lie along the positive real axis, and the UV ones, that
appear on the negative real axis. They are all double poles, with the
sole exception of the leading IR pole at $u=2$, related to the gluon
condensate, which is a simple one.

It will be important to consider the Borel transform of $\delta^{(0)}$ as well which reads~\cite{BJ08,BMO18}
\beq
B[\delta^{(0)}](u) = \frac{12}{(1-u)(3-u)(4-u)}\frac{\sin (\pi u)}{\pi u}B [\widehat D](u).\label{BorelDelta0}
\eeq
The analytic struture of this last Borel transform is much simpler
than that of $B[\widehat{D}](u)$. Now all the UV poles are simple
poles, because of the zeros of $\sin(\pi u)$.  For the same reason,
the leading IR pole of $B[\widehat D](u)$, at $u=2$, which is simple
in large-$\beta_0$, is cancelled in $B[\delta^{(0)}](u)$ --- a result
first pointed out in Ref.~\cite{BY92} for the Borel transformed spectral function. Our analysis with PAs benefits
greatly from these cancellations since the Borel transformed function
is now much less singular.\footnote{The fact that the only poles that remain double in Eq.~(\ref{BorelDelta0}) are the ones at $u=3$ and $u=4$ is not a coincidence. This reflects the fact that $\delta^{(0)}$ is maximally sensitive to the dimension-six and dimension-eight OPE condensates. Consequences of this general result for the choice of weight functions in $\alpha_s$ analyses
   from $\tau$ decays will be investigated elsewhere~\cite{inprep}.
} A simpler analytic structure can be much
more easily mimicked by the PAs. We also note that the leading UV pole
has a residue about ten times smaller than in the Adler function
counterpart. This, together with an enhancement of the residue of the double
pole at $u=3$, postpones the sign alternation of the series and enlarges the 
range of convergence of the Taylor series. PAs constructed to the expansion
of Eq.~(\ref{BorelDelta0}) benefit from these features of $B[\delta^{(0)}](u)$
and lead to smaller  errors by virtue of Pommerenke's theorem, granting
better coefficient's determination~\cite{Baker}.

The coefficients $c_{n,1}$ of the reduced Adler function can be reconstructed from the Borel transform by performing the expansion around $u=0$ and using  Eqs.~(\ref{eq:DCIPT}) and~(\ref{BorelDef}). The first six coefficients of the Adler function in the large-$\beta_0$ limit, denoted $\wh D_{L\beta}$, read ($N_f=3$, $\MSb$)
\beq
\wh D_{L\beta}(a_Q) = a_Q + 1.556 \, a_Q^2 + 15.71\,  a_Q^3 + 24.83\,  a_Q^4 + 787.8\,  a_Q^5 -1991  \, a_Q^6+ \cdots, \label{DLb}
\eeq
to be compared with their QCD counterparts given in Eq.~(\ref{eq:DCIPT}).  We observe that the sign alternation due to leading UV renormalon sets in at the sixth order (in the $\MSb$).
These coefficients lead to the following large-$\beta_0$ FOPT expansion of $\delta^{(0)}$:
\beq
\delta^{(0)}_{{\rm FO},L\beta}(a_Q) = a_Q + 5.119\,  a_Q^2 + 28.78\,  a_Q^3 + 156.7\,  a_Q^4 + 900.8\,  a_Q^5 +  4867 \, a_Q^6+\cdots,\label{FOPTLb}
 \eeq
to be compared with Eq.~(\ref{FOPTexp}). Now the sign alternation of
the coefficients is postponed and sets in only at the 9th order because
of the suppression of the leading UV pole in Eq.~(\ref{BorelDelta0}).
In comparison with the results in full QCD, the large-$\beta_0$ limit is
a good approximation, in the case of the Adler function, only up to
$\alpha_s^2$. However, for   $\delta^{(0)}_{{\rm FO},L\beta}$ this
approximation is still good up to the last known term,
i.e. $\alpha_s^4$. The reason for this better agreement lies in the
fact that these coefficients depend also on the $\beta$-function
coefficients --- which are largely dominated by $\beta_1$ in QCD ---
as well as on the integrals of Eq.~(\ref{FOPTdef}).

In Ref.~\cite{BMO18} we have performed a careful and systematic study
of the use of Pad\'e approximants to obtain the higher-order
coefficients of the series of Eqs.~(\ref{DLb}) and~(\ref{FOPTLb}). We
have verified that the procedure displays convergence and that the
leading renormalon poles can be correctly reproduced. We have also
discussed how renormalization scheme variations,   partial Pad\'e approximants~\cite{Baker}, as well as D-log Pad\'e approximants can be used in order
to improve the quality of the approximation. Finally, we  were able to
design an optimal strategy to predict the higher orders based only on
the first four coefficients of the series, which are the only ones
available in QCD. Here we will focus on the results from this strategy.

The optimal strategy of Ref.~\cite{BMO18} exploits the fact that the
Borel transform of $\delta_{\rm FO}^{(0)}$ displays a much simpler
singularity structure.  As shown in Eq.~(\ref{BorelDelta0}), this
Borel transform does not have the pole at $u=2$ and all other poles
are simple poles (with the exception of the ones at $u=3$ and
$u=4$). The leading UV renormalon is therefore more isolated from the
IR ones. It can be expected that the use of Pad\'e approximants
directly to this Borel transform should yield better and more stable
results than in the case of the Adler function. We should note that a
rational approximant to $\delta^{(0)}$ contains enough information to
allow for a full reconstruction of the Adler function since the  coefficients $c_{n,1}$ can easily be read off from the  FOPT
expansion of $\delta^{(0)}$ as
\begin{align}
  {\delta}^{(0)}_{\text{FO},L\beta} (a_Q) &= \  c_{1,1} \ a_Q + (3.563 \ c_{1,1} + c_{2,1}) \ a_Q^2 + (1.978 \ c_{1,1} + 7.125 \ c_{2,1}+c_{3,1}) \ a_Q^3  \nonumber \\
  & + (-45.31 \ c_{1,1} + 5.934 \ c_{2,1} + 10.69 \ c_{3,1} + c_{4,1}) \ a_Q^4+\cdots
\end{align}

\begin{figure}[!t]
\begin{center}
  \includegraphics[width=.51\columnwidth,angle=0]{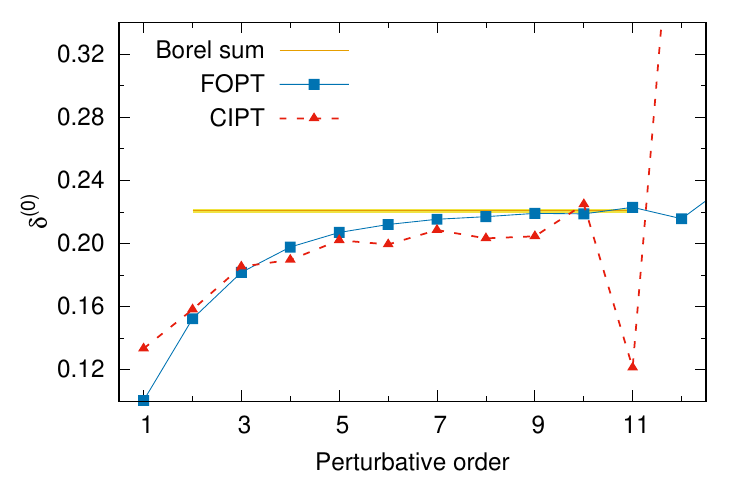}
  \includegraphics[width=.49\columnwidth,angle=0]{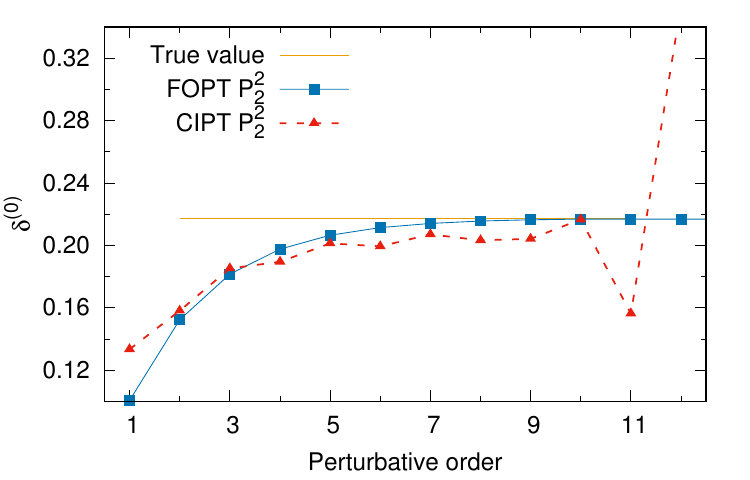}
  \includegraphics[width=.49\columnwidth,angle=0]{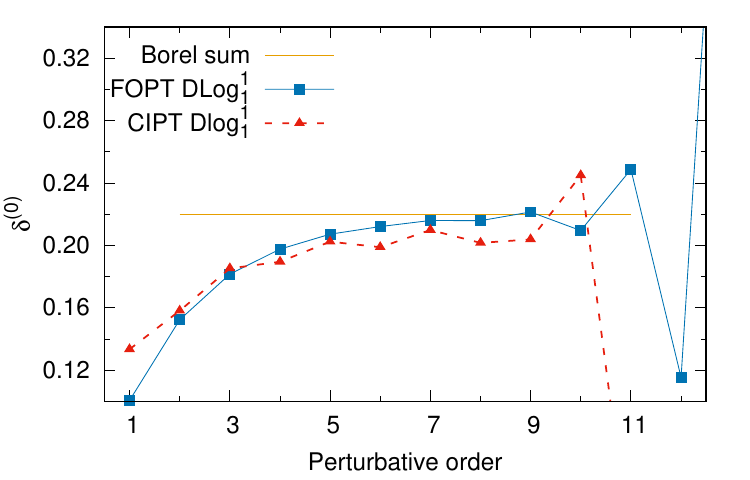}
\caption{Perturbative expansion of $\delta^{(0)}$ in FOPT and CIPT. (Upper panel)  exact large-$\beta_0$ limit, (lower-left panel)  results from  $P_2^2(a_Q)$, and (lower-right panel) results from ${\rm Dlog}_1^1(u)$. See text for the details regarding the approximants. 
}
\label{Fig1}
\end{center}
\end{figure}

We start here by applying Pad\'e approximants directly to the series
in $\alpha_s/\pi$, given by Eq.~(\ref{FOPTLb}).  As we have observed,
the FOPT series in large-$\beta_0$ is rather well behaved and, at
intermediate orders, its asymptotic nature is not visible yet. This is
mapped into a simpler analytic structure in the Borel plane.  It is
therefore likely that in this case the approximation of the series by
Pad\'e approximants in $a_Q$ will lead to a good description.
In Fig.~\ref{Fig1} (lower left panel)  we display an example of the results obtained (detailed numerical coefficients can be found in~\cite{BMO18}). The agreement with the exact results is quite impressive, as seen when comparing with the upper panel of Fig. \ref{Fig1}.

Another elegant and efficient way to obtain the higher-order coefficients
is to resort to   D-Log Pad\'es constructed to the Borel transform of $\delta_{\rm FO}^{(0)}$. This   turns out to be the
optimal way to improve the convergence while remaining completely
model independent. The success of this strategy can be understood from 
the  study of the function $F(u)=\frac{{\rm d}}{{\rm d}u}\log \left(B[\delta^{(0)}](u)   \right)$, introduced in Eq.~(\ref{Dlog}).
The leading analytic structure of $F(u)$  is much simpler than that of  the Adler function. The poles at $u=0$, $u=1$, and $u=2$ are exactly cancelled out by the presence of $\pi\cot(\pi u)$ leaving only a leading UV pole at $u=-1$, an IR pole at $u=3$ and a subleading IR pole at $u=4$.
It is therefore expected that the D-log Pad\'es should perform well in the present case, since the isolated simple poles can be reproduce by the rational approximant without the need of ``spending'' too many coefficients.

 In Fig.~\ref{Fig1} (lower-right panel) we present results for a D-Log Pad\'e applied to
$B[\delta^{(0)}]$. The predictions for $c_{5,1}$ have a rather  small
relative error and the sign
alternation is well reproduced by the Pad\'es using only the first  four coefficients
 as input. Their Borel integral provide excellent estimates for
the true value of the series. However, one must note that the results from the
D-Log Pad\'es applied to $B[\delta^{(0)}]$ are less good than those of
the Pad\'es applied to series in $\alpha_s/\pi$.  For example, for Dlog Pad\'es the coefficient $c_{4,1}$ is typically wrong
by a factor of about two while before it was only a few
percent off. Nevertheless, the description of the Borel transformed
$\delta^{(0)}$ by D-Log Pad\'es has the advantage that the factorial
growth of the coefficients is automatically reproduced and an asymptotic series is obtained,
in line with the exact result. Furthermore, Fig.~\ref{Fig1} shows that these small imperfections do not prevent an excellent reproduction
of the exact series.

In Fig.~\ref{Fig1} we  can see that both methods allow for an excellent reproduction  of the exact series up to around the 10th order. It is important to stress
that the superiority of FOPT, which is well established in large-$\beta_0$, is very well reproduced in both cases even
though we use as input only the first four coefficients of each series.

\section{Results in QCD}

In large-$\beta_0$  approximants constructed to
$\delta^{(0)}_{\rm FO}$ and $B[\delta^{(0)}] $ resulted optimal. Furthermore, for
the known terms of the perturbative series for $\delta^{(0)}_{\rm FO}$ in large-$\beta_0$ and
in QCD the coefficients rather similar. This suggests that the
regularity of the series is preserved in QCD, which indicates that it
can  be well approximated by Pad\'e approximants constructed directly
to the series in $\alpha_s/\pi$ as well. Moreover, although
Eq.~(\ref{BorelDelta0}) is strictly valid only in large-$\beta_0$,
because it relies on the one-loop running of the coupling,
modifications to this result would be solely due to higher-order beta function coefficients. We can therefore expect that a suppression of
the leading IR singularity at $u=2$, as well as a suppression of all
the other renormalons except for the ones at $u=3$ and $u=4$, would
survive in full QCD and render this Borel transform more amenable
to approximation by rational functions.

We start with Pad\'e approximants applied to the $\alpha_s/\pi$
expansion of $\delta^{(0)}_{\rm FO}$. We begin with a
post-diction of $c_{4,1}$ using $P_2^1(a_Q)$ and $P^2_1(a_Q)$. The
results for six higher-order coefficients obtained from these
approximants are shown in Tab.~\ref{Delta0PadesQCD}. The relative
error from the central values of $c_{4,1}$ is now only $\sim 13\%$. This is
quite remarkable when put into perspective since, before the true value
of $c_{4,1}$ was computed, a forecast of this coefficient using other
methods and including additional information (taking into account
known terms of order $\alpha_s^4 N_f^3$ and $\alpha_s^4N_f^2$) yielded
$c_{4,1}=27\pm 16$~\cite{BCK02,KS95,Kataev:1995vh}, a central value which was $45\%$
off. This gives an idea of how powerful optimal PAs can be.

\begin{table}[!t]
     \begin{center}{
  \caption{QCD Adler function coefficients  from  PAs constructed to the $\alpha_s$ expansion of  $\delta^{(0)}_{\rm FO}$. }
		\begin{tabular}{ccccccccc}
		\toprule
			& $c_{4,1}$ & $c_{5,1}$ & $c_{6,1}$ & $c_{7,1}$ &   $c_{8,1}$ & $c_{9,1}$ &  Pad\'e sum  \\
		\midrule
		$P^2_1$  &   $55.62$     &   $276.2$    &   $3865$      &   $1.952\times 10^4$   & $4.288\times 10^{5}$   & $1.289\times 10^{6}$  & 0.2080  \\
		$P^1_2$  &   $55.53$     &   $276.5$    &   $3855$      &   $1.959\times 10^4$   & $4.272\times 10^{5}$   & $1.307\times 10^{6}$  & 0.2079  \\
		$P^3_1$  &    input      &   $304.7$    &   $3171$      &   $2.442\times 10^4$   & $3.149\times 10^{5}$   & $2.633\times 10^{6}$  & 0.2053  \\
		$P^1_3$  &    input      &   $301.3$    &   $3189$      &   $2.391\times 10^4$   & $3.193\times 10^{5}$   & $2.521\times 10^{6}$  & 0.2051  \\
		\bottomrule
	\end{tabular}
     \label{Delta0PadesQCD}
     }
\end{center}
\end{table}

We turn then to the results obtained for $P_1^3$ and
$P_3^1$  which are also shown in
Tab.~\ref{Delta0PadesQCD}. Now, the forecasts of $c_{5,1}$ are 304.7 and 301.3, respectively. We note a striking stability of
the results for $c_{5,1}$ and $c_{6,1}$; even $c_{7,1}$ and $c_{8,1}$
are remarkably similar in all of the four approximants considered.
The use of the PAs to sum the asymptotic series also leads to consistent
result in  all cases, as can be
seen in the last column of Tab.~\ref{Delta0PadesQCD}.
Our experience from large-$\beta_0$ indicates that this stability and the good
prediction of $c_{4,1}$ strongly corroborate the robustness of the results.
We have checked that the use of D-log Pad\'e approximants is also very successful. 
We are, therefore, in a position to conclude that using PAs to
$\delta^{(0)}_{\rm FO}$ in QCD is at least as stable as in
large-$\beta_0$. We should then investigate the approximants
constructed to its Borel transformed.

As in the previous section, the quality of the forecast of $c_{4,1}$
as well as stability arguments lead us to conclude that the D-log
Pad\'es are the optimal approximants to
$B[\delta^{(0)}](u)$. Higher-order coefficients obtained from D-log
Pad\'es constructed to $B[\delta^{(0)}](u)$ in QCD are shown in
Tab.~\ref{Delta0BorelPadesQCD}. Now, the postdiction of the last known
coefficient, $c_{4,1}$, has a relative error of only about $\sim 6\%$,
about half of what was obtained with Pad\'es to the series in
$\alpha_s$. Also, the stability of the results when using the exact
value of $c_{4,1}$ as input is quite remarkable. The results for
$c_{5,1}$ and $c_{6,1}$ are rather stable not only among the D-log
Pad\'es of Tab.~\ref{Delta0BorelPadesQCD} but also when compared with
the results of Tab.~\ref{Delta0PadesQCD}. The   approximant,
$\Dlog^1_1$, not shown in Tab.~\ref{Delta0BorelPadesQCD}, leads to slightly lower values for the coefficients (e.g., $c_{5,1}= 237$), but
even these apparent instability can  be well understood in terms of a partial cancelation between a pole and a  zero present in the $P_1^1$
used for its construction~\cite{Baker}. We, therefore, consistently discard this approximant. 
 It
is also interesting to observe that all the D-log Pad\'es of
Tab.~\ref{Delta0BorelPadesQCD} predict that the sign alternation of
the series starts at order 11.  This suggests that the UV singularity in QCD is less prominent 
than in large-$\beta_0$ which
should postpone the sign alternation, a fact that can be corroborate by 
scheme
variations of the type of~\cite{BJM16} as discussed in detail in~\cite{BMO18}.
  Finally, the Borel sum of the
series obtained from these D-log Pad\'es is also very consistent (last
column of Tab.~\ref{Delta0BorelPadesQCD}).

\begin{table}[!t]
   \begin{center}{
  \caption{QCD Adler function coefficients  from D-Log Pad\'e approximants   to  $B[\delta^{(0)}](u)$. }
	\begin{tabular}{ccccccccc}
		\toprule
		& $c_{4,1}$ & $c_{5,1}$ & $c_{6,1}$ & $c_{7,1}$ &   $c_{8,1}$ & $c_{9,1}$ & Borel sum\\
		\midrule 
		$\text{DLog}^1_0$  &   $51.90$     &   $272.6$    &   $3530$     &   $1.939\times 10^4$    & $3.816\times 10^{5}$   & $1.439\times 10^{6}$  & 0.2050  \\
		$\text{DLog}^0_1$  &   $52.08$     &   $273.7$    &   $3548$     &   $1.953\times 10^4$    & $3.840\times 10^{5}$   & $1.456\times 10^{6}$  & 0.2052  \\
		$\text{DLog}^2_0$  &    input      &   $254.1$    &   $3243$     &   $1.725\times 10^4$    & $3.447\times 10^{5}$   & $1.187\times 10^{6}$  & 0.2012 \\
		$\text{DLog}^0_2$  &    input      &   $256.4$    &   $3271$     &   $1.769\times 10^4$    & $3.493\times 10^{5}$   & $1.258\times 10^{6}$  & 0.2019 \\
		\bottomrule
	\end{tabular}
     \label{Delta0BorelPadesQCD}
     }
\end{center}
\end{table}

The picture that emerges from the results of this section is that the
use of $\delta^{(0)}_{\rm FO}$ and its Borel transform lead to the best model-independent
approximants in QCD --- as is the case in large-$\beta_0$.  The
quality of the predictions of $c_{4,1}$ as well as the stability of
the results among different approximants signal that we have managed
to obtain a robust description of $\delta^{(0)}$ and of the Adler function at higher
orders.

We extract our final estimates for the higher-order
coefficients from the eight approximants of Tabs.~\ref{Delta0PadesQCD}
and~\ref{Delta0BorelPadesQCD} including, thus, those that have only
three coefficients as input parameters. By doing so, we take advantage
of Pad\'es that belong to different sequences and can obtain a more
reliable error estimate for our final coefficients. Since one of the
most striking features of these results is their stability, we will
not try to favour one approximant over another, even though one could
try to inspect their analytic structure in detail with this goal in mind.  Our
final estimate of the coefficients and of the true value of
$\delta^{(0)}$ is obtained as the average of the eight results of
Tabs. \ref{Delta0PadesQCD} and~\ref{Delta0BorelPadesQCD}.
To these averages we add an error equal to the maximum spread
found between the coefficients obtained from two different approximants.
This error should certainly not be interpreted in a statistical sense; it
gives an interval  where the value of the coefficient is expected to lie.

This procedure applied to the six-loop coefficient, $c_{5,1}$, leads to
\beq
c_{5,1} =  277 \pm 51,\label{finalc51}
\eeq
which largely covers all the results obtained from our optimal
approximants.  Therefore, in a sense, our error estimate could even be
considered as too conservative --- even if much smaller than other
estimates in the literature. For example, in Ref.~\cite{BJ08} the estimate $c_{5,1}
=283 \pm 142$ is used, while in Ref.~\cite{BCK02} one finds $c_{5,1}=
145\pm 100$ (using only partial information about the five-loop coefficient). The value obtained from the principle of Fastest Apparent Convergence (FAC) in Ref.~\cite{BCK08} is $c_{5,1}=275$, remarkably close to our final central value, given in Eq.~(\ref{finalc51}).
On the basis of what we know about the series
  coefficients, it seems extremely unlikely that the six-loop
  coefficient would not be within these bounds.

Results for coefficients $c_{6,1}$ and higher are given in
Tab.~\ref{FinalCoeff}. The final values for the Adler function
coefficients are extracted with reasonable errors up to
$c_{10,1}$. One should remark that due to the $\alpha_s$ suppression
at these higher orders, an error that seems large in the coefficient
does not translate into a very large uncertainty in the sum of the
series. The situation changes only for $c_{11,1}$. For this
coefficient, six of the PAs of Tabs.~\ref{Delta0PadesQCD}
and~\ref{Delta0BorelPadesQCD} predict that the sign alternation  sets in. However, two of the approximants do not, which
leads to the huge error. Therefore, we find some indication that the
sign alternation of the Adler function coefficients sets in at the
eleventh order (in agreement with \cite{BJ08}). This instability signals that our results cease to be
fully reliable at the  11th order.

\begin{table}[!t]
\begin{center}
  {
     \caption{Final values for the QCD Adler function coefficients  obtained from  PAs to  $\delta^{(0)}_{\rm FO}$. }
	\begin{tabular}{cccc}
		\toprule
			 $c_{5,1}$ & $c_{6,1}$ & $c_{7,1}$ &   $c_{8,1}$   \\
		       $277\pm 51$   &   $3460\pm 690$    &  $(2.02 \pm 0.72)\times10^4$    &   $(3.7\pm 1.1)\times 10^5$   \\
		\midrule 
			 $c_{9,1}$ & $c_{10,1}$ & $c_{11,1}$ &   $c_{12,1}$   \\
		       $(1.6\pm 1.4)\times10^6$   &   $(6.6\pm 3.2)\times10^7$    &  $(-5\pm 57)\times10^7$    &   $(2.1\pm 1.5)\times 10^{10}$   \\
		\bottomrule
	\end{tabular}\label{FinalCoeff}
}
\end{center}
\end{table} 

We apply the same procedure described above to obtain an estimate for
the true value of the $\delta^{(0)}$ using the results in the last
columns of Tabs.~\ref{Delta0PadesQCD} and~\ref{Delta0BorelPadesQCD}.
Using $\alpha_s(m_\tau^2) = 0.316 \pm 0.010$~\cite{PDG18}, this leads to
\beq
\delta^{(0)} = 0.2050 \pm 0.0067 \pm 0.0130,
\label{delta0Final}
\eeq
where the first error is the estimate from the spread of the PAs and the
second error is due to the uncertainty  in $\alpha_s$. This result agrees
with other estimates found in the literature using other 
methods~\cite{BJ08,BJM16,delta0PMC,delta0conformalC})

\begin{figure}[!t]
\begin{center}
  \includegraphics[width=.7\columnwidth,angle=0]{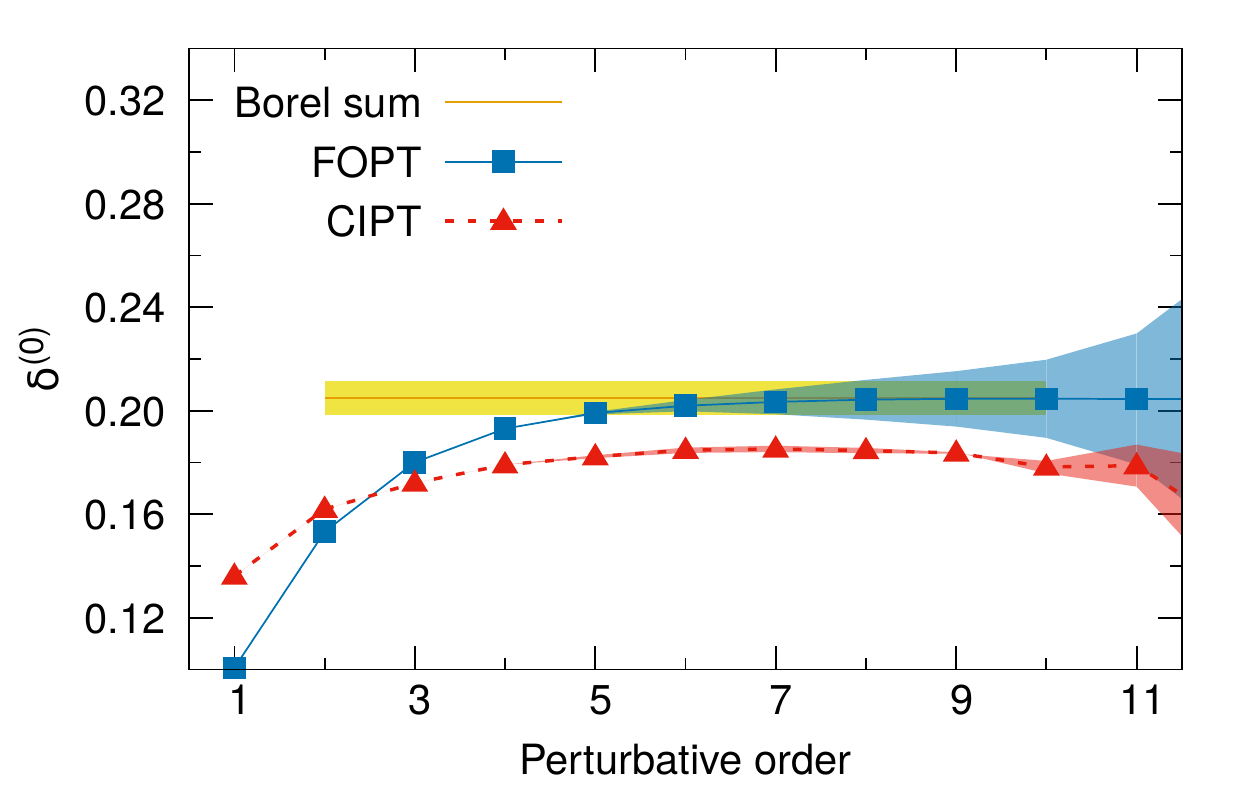}
\caption{Final results for  $\delta^{(0)}$ in QCD using the coefficients of Tab.~\ref{FinalCoeff} and the result of Eq.~(\ref{delta0Final}). The bands in the perturbative expansions reflect the uncertainty in the coefficients while the band in the sum of the series is obtained from the spread of the values from individual PAs (last columns of Tabs.~\ref{Delta0PadesQCD} and~\ref{Delta0BorelPadesQCD}). We use $\alpha_s(m_\tau^2) = 0.316.$ }
\label{delta0FinalPlot}
\end{center}
\end{figure}

With the coefficients of Tab.~\ref{FinalCoeff} we are finally in a
position to plot, in Fig.~\ref{delta0FinalPlot}, the perturbative
expansions of $\delta^{(0)}$ and compare them with the true value of the
series obtained from Eq.~(\ref{delta0Final}). The bands in the
perturbative expansions of Fig.~\ref{delta0FinalPlot} represent the
uncertainty from the series coefficients, given in
Tab.~\ref{FinalCoeff}, while the band in the Borel sum of the series
is the first error Eq.~(\ref{delta0Final}).
The uncertainties we are able to obtain from the optimal Pad\'e approximants
allow us to conclude that FOPT is the favored renormalization-scale setting
procedure  in the case of QCD. The CIPT series, even though it looks
more stable around the fourth order, does not approach well the central value of the
sum of the series. The recommendation that FOPT is the best procedure in QCD
was advocated in Ref.~\cite{BJ08} in the renormalon-model context. Here it is
reobtained  in a  model-independent way.

\section{Conclusion}

In this work we have used the mathematical method of Pad\'e approximants to obtain a description of the perturbative
QCD series for hadronic $\tau$ decays  beyond five loops. We have discussed strategies to optimize the use of the available knowledge --- namely the first four coefficients. The Borel transform of the series
can be used to explain why these strategies are so efficient, as can be cross-checked from the exact results in the large-$\beta_0$ limit. 
The method is shown to provide accurate and reliable predictions for the higher orders and for the sum of the series. This can then be used
to study the problem of renormalization-scale setting in hadronic $\tau$ decays and the result of this analysis is that fixed-order perturbation theory, FOPT,
is favoured within our model-independent reconstruction of the series. 

Perturbative expansions in QCD, such as the one for  hadronic $\tau$ decays, are divergent series that are assumed to be asymptotic. Any conclusion about the renormalization group improvement of the series must be drawn in this context, which automatically invalidades arguments based on the ``convergence" of 
the different series. Since a few years, there is solid renormalon-based evidence that FOPT is the best method to set the scale in $\tau$ decays~\cite{BJ08,BBJ13}. In this work, we have used a completely model-independent method, namely the Pad\'e approximants, to reconstruct the higher orders~\cite{BMO18}. Our final results are rather similar to the ones obtained from the renormalon-based methods and fully corroborate the conclusions of Refs.~\cite{BJ08,BBJ13}. Therefore, as of 2018, the evidence in favour of FOPT is significant and makes this procedure, most likely, the  best one to be used in phenomenological studies of hadronic $\tau$ decays.


\paragraph{Funding information}
This work was
partially supported by the S\~ao Paulo Research Foundation (FAPESP)
grants 2015/20689-9 and 2016/01341-4 and by CNPq grant number 305431/2015-3.
The work of P.M. is supported by the Beatriu de Pin\'os postdoctoral
programme of the Government of Catalonia's Secretariat for
Universities and Research of the Ministry of Economy and Knowledge of
Spain.



\bibliography{SciPost_Example_BiBTeX_File.bib}

\end{document}